\documentstyle[11pt,newpasp,epsfig,twoside]{article}
\def\edcomment#1{\iffalse\marginpar{\raggedright\sl#1\/}\else\relax\fi}
\reversemarginpar

\begin{document}
\title{Spiral Galaxy - ICM Interactions in the Virgo Cluster}
 \author{Jeffrey D. P. Kenney}
\affil{Yale University}
\author{Hugh Crowl}
\affil{Yale University}
\author{Jacqueline van Gorkom}
\affil{Columbia University}
\author{Bernd Vollmer}
\affil{Observatoire Astronomique de Strasbourg}

\begin{abstract}
We discuss HI and optical evidence for ongoing ICM-ISM interactions in
6 HI-deficient Virgo cluster spiral galaxies.
One of the clearest cases is the highly inclined Virgo galaxy NGC~4522,
which has a normal stellar disk but a truncated gas disk,
and lots of extraplanar gas right next to the gas truncation radius in the disk.
Unusually strong HI, H$\alpha$ and radio continuum emission are all 
detected from the extraplanar gas. The radio continuum polarized
flux and spectral index peak on the side opposite the extraplanar gas,
suggesting ongoing pressure by the ICM.
Four other HI-deficient edge-on Virgo spirals 
show evidence of extraplanar ISM gas or
exhibit asymmetries in their disk HI distributions,
but contain much less extraplanar HI than NGC~4522.
Comparison with recent simulations suggests this difference may be
evolutionary, with large surface densities of extraplanar gas observed 
only in early phases of an ICM-ISM interaction.
In NGC~4569, the H$\alpha$ image shows 2 effects of ICM pressure on the galaxy ISM.
An anomalous arm of HII regions, possibly extraplanar, emerges from the edge of
a truncated H$\alpha$ disk. This resembles the arms seen in simulations which are
formed by the combined effects of wind pressure plus rotation.
An extended nebulosity near the minor axis, also in the NW,
is interpreted as a starburst outflow bubble disturbed by ICM wind pressure.

\end{abstract}

\section{Introduction}

Interactions between the interstellar medium (ISM) of a galaxy and
the gas in the intracluster medium (ICM) are believed to be
one of the main processes which drive galaxy evolution in clusters
(Gunn \& Gott 1972; Poggianti et al 1999; Koopmann \& Kenney 2003; van Gorkom 2003).
Gas stripping is likely responsible for the transformation of many cluster
spirals into  S0's and Sa's, and thus may help cause both the morphology-density
relation (Dressler et al 1997)
and the Butcher-Oemler effect (Butcher \& Oemler 1978).
ISM stripping is also significant for the evolution of the ICM, as
gas removed from the galaxies enriches the ICM with heavy elements.

There is widespread evidence for ICM-ISM stripping in cluster spiral galaxies,
from single dish HI studies showing that many cluster galaxies are
HI-deficient (e.g., Solanes et al 2002) ,
and interferometer HI studies showing many cluster spirals
with truncated gas disks (Cayatte et al 1990; Bravo-Alfaro et al 2001).
Since a galaxy will be deficient in gas and have a truncated gas disk
long after an interaction, it is not
clear whether stripping is active or occurred in the past for most cluster galaxies.
Here we discuss HI and optical evidence for ongoing ICM-ISM interactions in
6 HI-deficient cluster spiral galaxies in the nearby Virgo cluster (16 Mpc).
With detailed observations of galaxies experiencing ICM-ISM stripping
we can begin to learn what actually happens in an ICM-ISM interaction,
and the consequences for galaxy evolution.

\section{NGC~4522}

\begin{figure}[h]
\plottwo{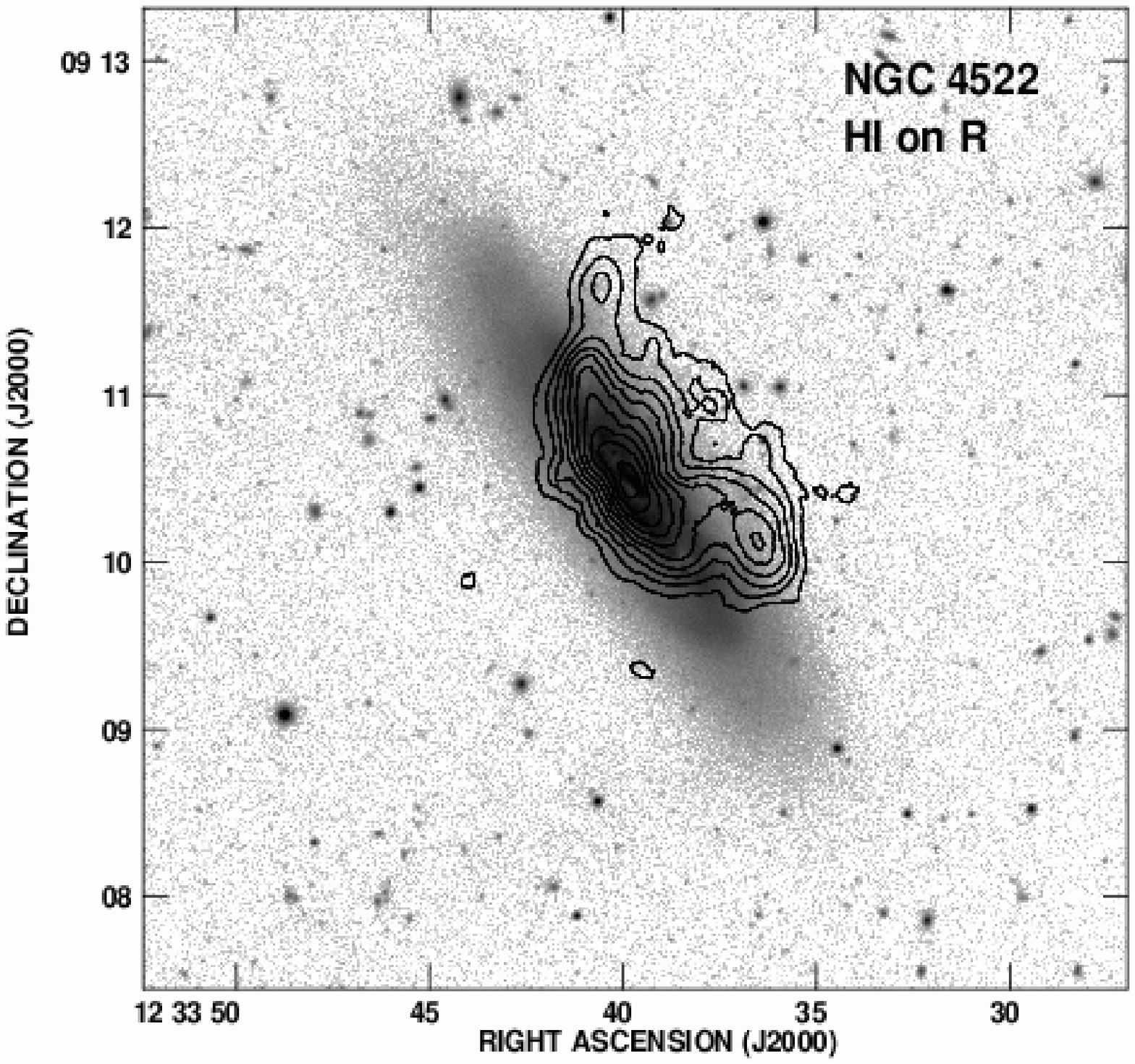}{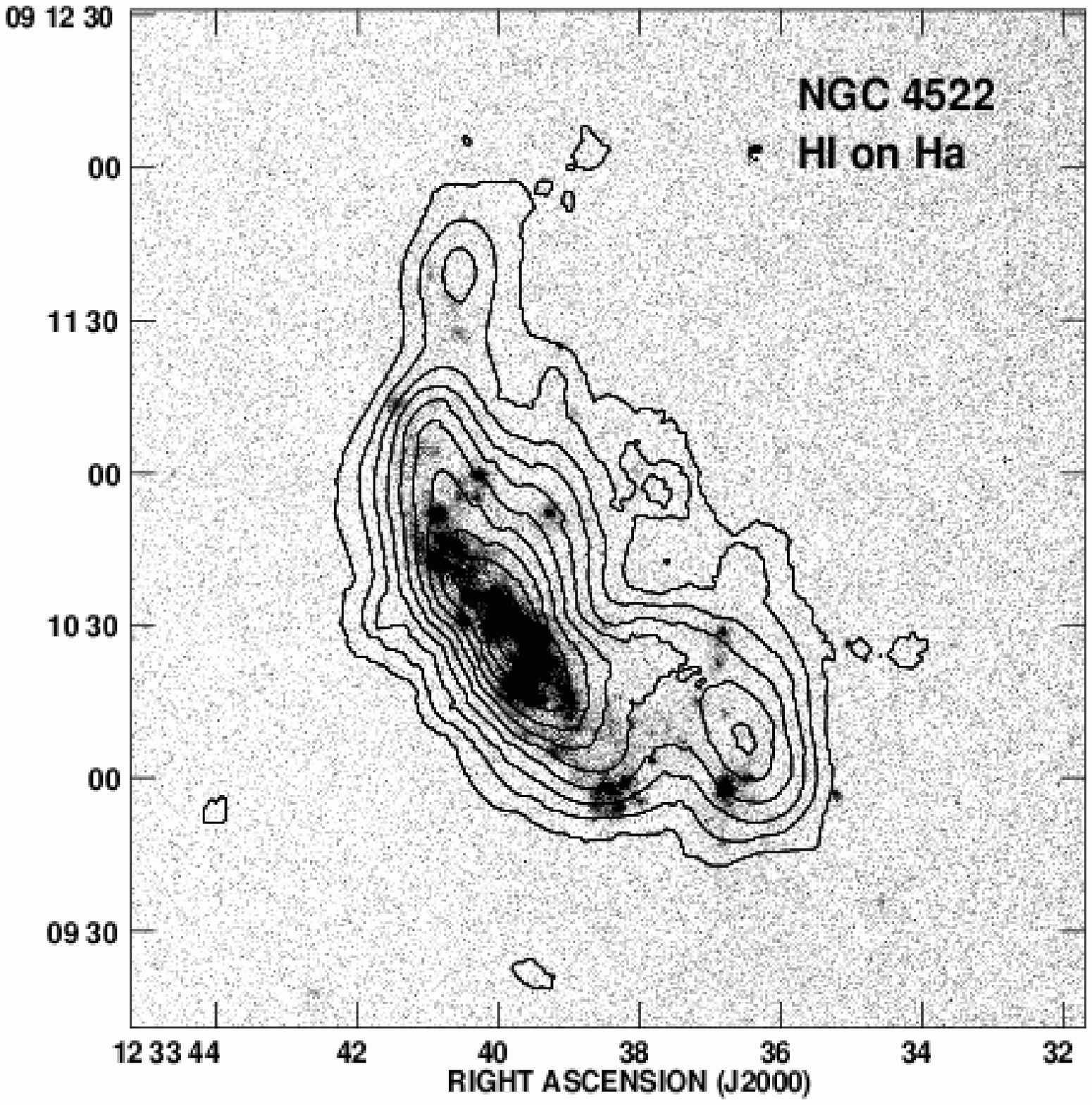}
\caption{
a.) VLA HI total intensity contour map of NGC~4522 on WIYN R-band greyscale image.
Lowest HI contour level and contour increments are 1.6 M$_{\sun}$ pc$^{-2}$.
Note undisturbed outer stellar disk. 
b.) VLA HI contour map on H$\alpha$ greyscale image of NGC~4522.
HI map from Kenney et al (2003) and optical images from Kenney \& Koopmann (1999).}
\end{figure}

One of the clearest and nearest
cases of a cluster spiral with an ongoing ICM-ISM interaction
is the highly inclined Virgo L* galaxy NGC~4522.
Figure 1 shows the VLA HI map of NGC~4522 on an optical image
(Kenney et al 2003; Kenney \& Koopmann 1999),
revealing an undisturbed stellar disk and a selectively disturbed ISM.
The gas disk, including HI, H$\alpha$
and radio continuum emission, is truncated beyond 0.35R$_{25}$.
A large quantity of extraplanar gas is detected on one side of the disk,
close to the disk truncation radius.

There are indications from the HI morphology and kinematics,
and from the radio continuum data, that
the galaxy is experiencing active ICM pressure.
On the side of the galaxy rotating into the  oncoming ICM (the SW side),
which is expected to experience stronger pressure,
there is much less disk HI and more extraplanar HI.
Also, much of extraplanar gas has a modest blueshift with respect to nearby disk emission, 
and some of the extraplanar gas has large linewidths with FWZI=150 km s$^{-1}$,
suggesting acceleration toward the mean cluster velocity.
VLA radio continuum maps show that both
the polarized flux and the spectral index both peak to the east of the
nucleus, on the side of the galaxy opposite the extraplanar gas
(Vollmer et al 2003).
This suggests ISM compression and particle acceleration on the eastern
leading edge of the galaxy.

NGC~4522 is 3.3$\deg$ from M87, far enough from the cluster core
that the standard ICM ram pressure 
seems insufficient by an order of magnitude to produce the observed stripping.
This suggests that either we are observing the galaxy in a post-peak-pressure phase, 
or that the ICM pressure on NGC~4522
is signficantly stronger than is commonly assumed in a static, homogenous ICM.
Peak ram pressures could easily be an order of magnitude higher
in a shock-filled dynamic ICM with large scale flows (Dupke \& Bregman 2001).

\section{Other Highly-Inclined HI-Deficient Spirals}

We have recently observed HI with VLA in C-array
in 4 Virgo spirals chosen to be similar to NGC~4522 (Crowl et al in prep).
These galaxies, including NGC~4522, 
all have similar optical luminosity (within a factor of 2),
are all highly inclined ($\geq$65 deg),
are all $\sim$2-3.5 degrees from M87,
and are HI deficient by factors of 4-10.

Figure 2 shows the VLA HI maps on R band optical images of the galaxies.
All have strongly truncated HI disks.
All show HI asymmetries in the disk, which is evidence of an ongoing interaction.
IC~3392, NGC~4402, and possibly NGC~4419 have weak extraplanar HI emission,
but it is an order of magnitude weaker than in NGC~4522.
Extraplanar HI has not been detected in the VLA maps of NGC~4388,
but there is a single dish detection of extraplanar HI toward the NE (Vollmer \& Hutchmeier 2003)
The galaxy also has extraplanar optical emission line regions, which may be stripped gas
ionized by the active nucleus (Yoshida et al 2003).
NGC~4402 also has extraplanar radio continuum emission and disturbed HI kinematics in the disk.

\begin{figure}[h]
\plotone{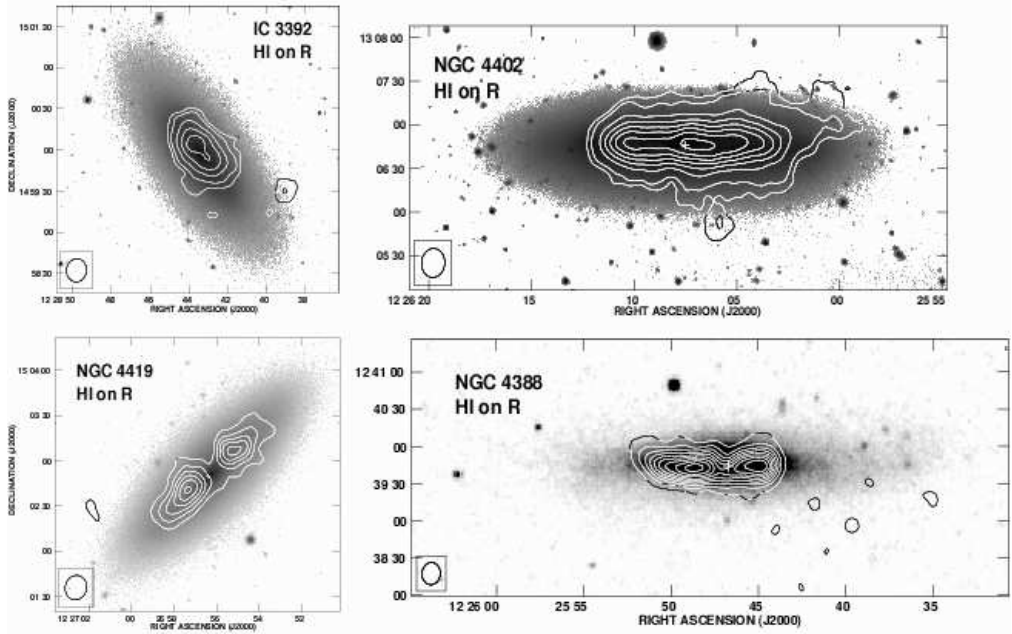}
\caption{VLA HI maps on optical images of Virgo spirals IC~3392, NGC~4402, NGC~4419 and NGC 4388.
NGC~4388 and NGC~4419 both have central absorption of HI due to strong nuclear continuum sources.
 From Crowl et al (2004).}
\end{figure}

None of the 4 galaxies has such strong extraplanar HI emission as NGC~4522.
Whereas the peak extraplanar HI in NGC~4522 has a surface density of 10 M$_{\sun}$ pc $^{-2}$,
and is 70\% of the maximum disk surface density,
the other galaxies have extraplanar peaks with surface densities of $\leq$1 M$_{\sun}$ pc $^{-2}$,
and $\leq$20\% of the maximum disk surface density.
What makes NGC~4522 different from the other 4 galaxies?
One possibility is that NGC~4522 is in a different evolutionary phase of stripping than
the other galaxies.
Only at early times in a stripping event do
simulations show a large concentration of extraplanar gas
near the disk truncation radius (Schulz \& Struck 2001; Vollmer et al 2001).
At later times (indeed at most times), the gas is more diffuse.

Four of the 5 have high line-of-sight cluster velocities
(V$_{\rm gal}$-V$_{\rm cl}$$>$1000 km s$^{-1}$),
so that their motion must be mostly along the line-of-sight,
with smaller components in the plane of the sky.
In these cases, the likelihood is high that we know 
which side is rotating into the ICM wind.
In NGC~4522 there is less disk HI on the side rotating into the ICM,
but more in the halo.
However, in NGC~4388, NGC~4419 and NGC~4402, 
there is more disk HI on the side rotating into the ICM,
and relatively little HI in the halo.

NGC~4388 and NGC~4402 are within 2$\deg$ of M87, and even closer to M86,
which may be part of a merging sub-cluster (Schindler et al 1999).
Most galaxies near M86 seem to have ongoing interactions.
At a distance of 3$\deg$ from M87, and
with a weak extraplanar gas arm and a symmetric H$\alpha$ ring at the truncation radius,
IC~3392 may well be in a post-peak pressure phase.
IC~3392 and NGC~4580 (H$\alpha$ images shown in Koopmann et al 2001) have bright rings of star
formation located at the truncation radius, and may be examples of ``annealed'' disks.
Annealed disks have enhancements in gas surface density at the edge of truncated gas disk,
and are produced in later stages of ICM-ISM interaction simulations
(Schulz \& Struck 2001).

\section{NGC~4569}

\begin{figure}[h]
\plottwo{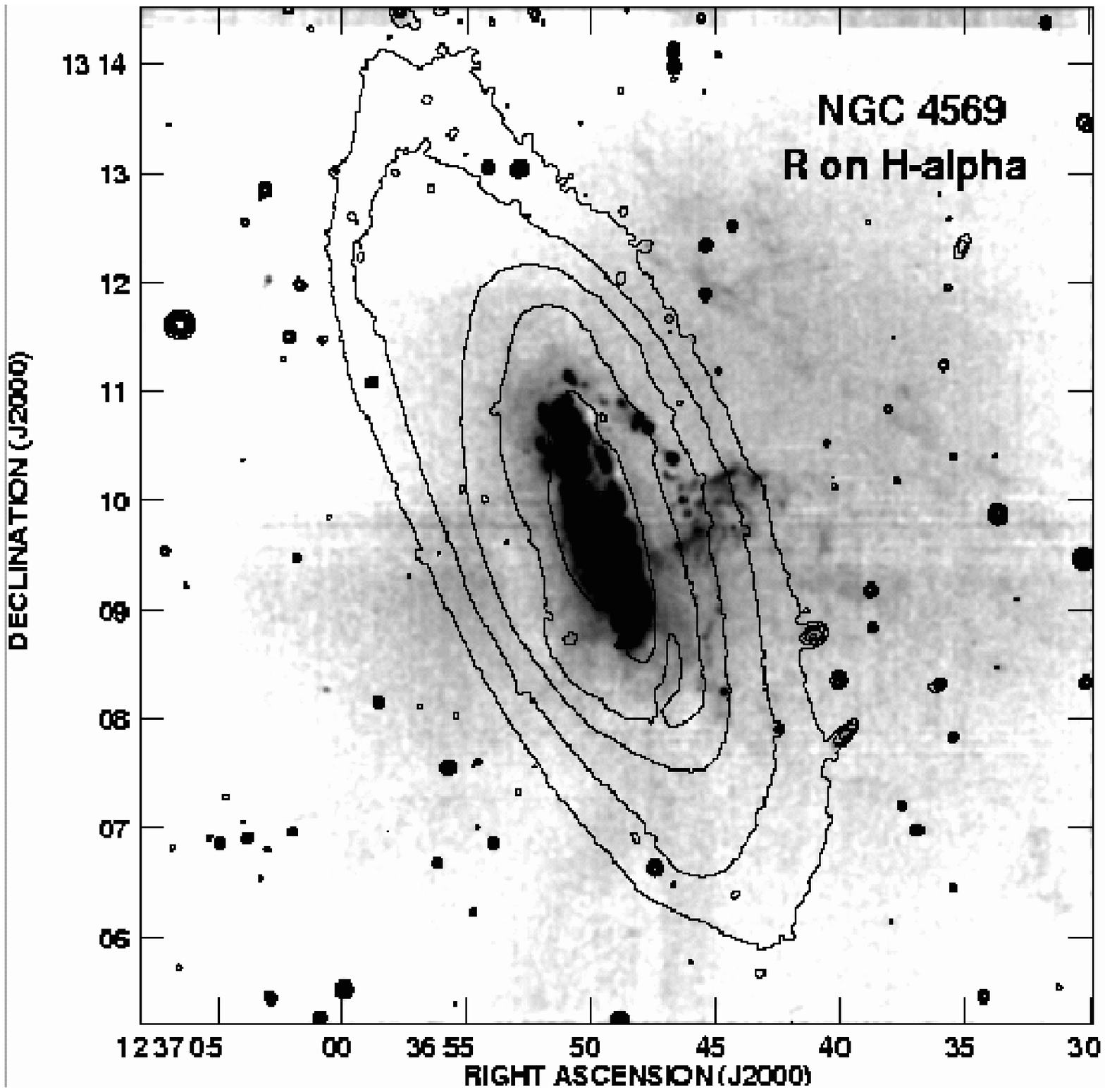}{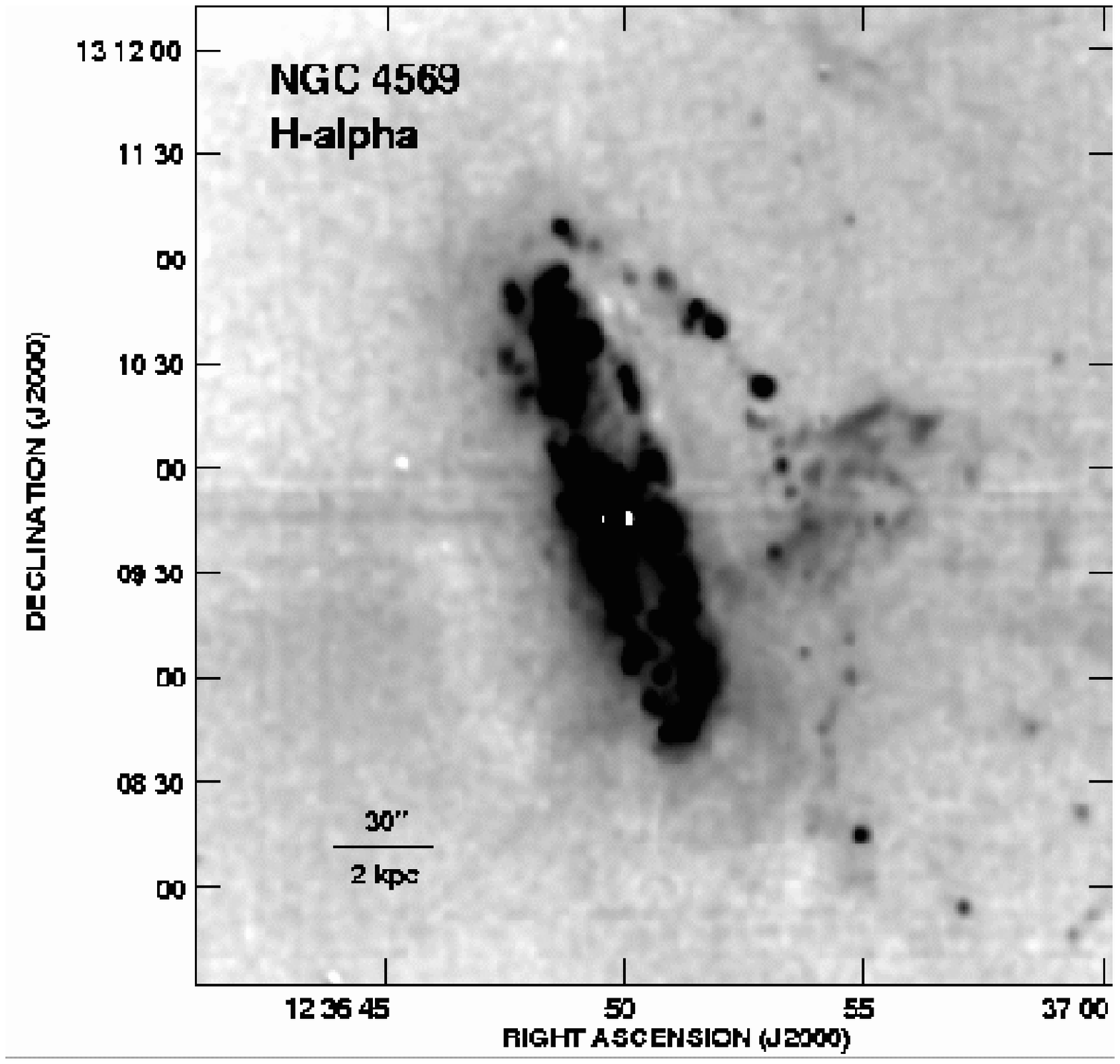}
\caption{H$\alpha$ image of NGC~4569,
from a WIYN image convolved to a resolution of 3$''$
shown with R-band contours in the left panel.
H$\alpha$ emission is truncated at a radius of 80$''$=6 kpc, and
is absent from the outer disk. An anomalous arm of HII regions 
which may be gas stripped from the disk
emerges from the northern end of the truncated gas disk. 
The diffuse nebulosity $\sim$1$'$ west of the nucleus
may be a starburst outflow bubble disrupted by ICM pressure.
}
\end{figure}

NGC~4569 (M90) is among the brighter spiral galaxies in the Virgo cluster, and
1-2 magnitudes brighter than the other 5 spirals discussed in this paper.
It is close (1.9$\deg$) to the cluster center, and
is also one of the few blueshifted  galaxies on the sky (v= -235 km/s), moving
through the cluster with a very high line of sight velocity of -1300 km/s.
Thus, its orbit likely has a large radial component, much of which is directed toward us.  
It has one tenth the HI content of a typical isolated galaxy of its size (Giovanelli \& Haynes 1983),
and a truncated HI disk (Warmels 1988; Cayatte et al 1990).
NGC~4569 and NGC~4522 are the 2 Virgo spirals with numerous 
extraplanar HII regions detected in the Koopmann et al (2001) Virgo survey.

An H$\alpha$ image of NGC~4569 taken with the Mini-Mosaic camera on the WIYN telescope
(Figure 3) shows 2 effects of ICM pressure on the galaxy ISM
(Kenney \& Hameed, in prep): an anomalous arm of HII regions, and a diffuse
nebulosity near the minor axis.
Both features are also described but interpreted differently by Tschoke et al (2002).

The H$\alpha$ image shows a star forming disk sharply truncated at 30\% of the optical radius.
There is virtually no  H$\alpha$ emission from the outer disk.
An anomalous arm of HII regions emerges from the northern end of the truncated gas disk, and extends
for over 3$'$=13 kpc toward the SW.
It resembles the arms
that are seen in some phases of ICM-ISM interaction simulations (Vollmer et al 2001; Schulz \& Struck 2001),
in which much of the gas stripped from the disk forms one dominant extraplanar arm,
due to the combined effects of wind pressure plus galaxy rotation.

The H$\alpha$ image also shows a faint, extended, diffuse nebulosity near the NW minor axis, 
on the same side of the major axis as the  anomalous arm.
There is no evidence of an eastern counterpart in the optical, although something
a few times fainter than the western nebulosity would not be detected with the present data.
The nebulosity is displaced to the S of the minor axis,
and its highest surface brightness features form a relatively sharp northern boundary.
Both the nebular morphology and the anomalous arm location are suggestive of
ICM pressure from the N or NE.
Optical spectroscopy indicates a nuclear starburst with an age of 5-6 Myr (Gabel \& Bruhweiler 2002).
Thus we suspect the optical nebulosity is a starburst outflow bubble,
partially disrupted by ICM pressure.

\acknowledgments{This research is partially funded by grant AST-0071251 from the National Science Foundation.}

\end{document}